# Machine learning prediction of the MJO extends beyond one month

Tamaki Suematsu[1]*, Kengo Nakai[2]*, Tsuyoshi Yoneda[3], Daisuke Takasuka[4,5], Takuya Jinno[6,3], Yoshitaka Saiki[7], Hiroaki Miura[6]

[1]RIKEN Center for Computational Science; Kobe, Hyogo, 650-0047, Japan.

*Tamaki Suematsu. E-mail: tamaki.suematsu@riken.jp

[2]Faculty of Marine Technology, Tokyo University of Marine Science and Technology; Tokyo 135-8533, Japan.

*Kengo Nakai. E-mail: knakai0@kaiyodai.ac.jp

[3]Graduate School of Economics, Hitotsubashi University; Kunitachi, Tokyo, 186-8601; Japan.

[4]Atmosphere and Ocean Research Institute, The University of Tokyo; Kashiwa, Chiba, 277-0882, Japan.

[5]Japan Agency for Marine-Earth Science and Technology; Yokohama, Kanagawa, 236-0001, Japan.

[6]Graduate School of Science, The University of Tokyo; Bunkyo-ku, Tokyo, 113-0033, Japan.

[7]Graduate School of Business Administration, Hitotsubashi University; Kunitachi, Tokyo, 186-8601, Japan.

*These authors contributed equally to this work

## Abstract

The prediction of the Madden-Julian Oscillation (MJO), a massive tropical weather event with vast global socio-economic impacts[1,2], has been infamously difficult with physics-based weather prediction models[3–5]. Here we construct a machine learning model using reservoir computing technique that forecasts the real-time multivariate MJO index (RMM)[6], a macroscopic variable that represents the state of the MJO. The training data was refined by developing a novel filter that extracts the recurrency of MJO signals from the raw atmospheric data and selecting a suitable time-delay coordinate of the RMM. The model demonstrated the skill to forecast the state of MJO events for a month from the pre-developmental stages. Best-performing cases predicted the RMM sequence over two months, which exceeds the expected inherent predictability limit of the MJO.

**Main text**

The Madden–Julian Oscillation (MJO) [7] is a massive cluster of convective activities in the tropics that spans thousands of kilometers traveling slowly eastward from the Indian Ocean to the central Pacific in approximately 20 to 60 days. It has far reaching influence on the global weather [1,8] and is recognized to be one of the most important sources of predictability in extended-range weather forecast longer than weeks [9,10]. However, simulation of the MJO by physics-based dynamical numerical models (hereafter dynamical models) has been shown to be notoriously difficult [3,11,12]. It has only been since the mid 2000s that the predictability of the MJO by dynamical models [13,14] exceeded that of empirical statistical models, such as atmospheric-only linear inverse models, at two to three weeks [15]. The current forecast skills of dynamical models for MJO prediction lie between two to five weeks [16], which falls short of the expected inherent predictability of the MJO estimated from multi-model ensemble studies at six to seven weeks [17].

Weather forecasts by dynamical models require physical parameterizations that incorporate the mean effects of the sub-grid scale processes on the evolution of the grid-scale flows. However, parameterizations are prone to empirical tuning and are inevitable sources of uncertainty of the dynamical models [18–20] because we have yet to determine the correct theoretical formulations and parameters of the statistical mean states of microscopic processes. The difficulties of reducing the ambiguities in parameterizations continue to be a major limiting factor for improving dynamical models and for identifying processes essential for successful weather predictions. In contrast, machine learning models with relatively small number of neural networks trained only by the time series of macroscopic variables has the potential to implicitly incorporate the influence of microscopic variables on the macroscopic variables and to eliminate the parameterizations that unsatisfactorily replicate the multiscale interactions between the unresolved and the resolved processes.

The effectiveness of the machine learning methods has been demonstrated in the fields of atmospheric and climate science. Considerable progress has been achieved in areas for forecasting phenomena with large socio-economic impacts such as the El Niño Southern Oscillation [21,22], Asian summer monsoons [23], and hurricanes [24] at incomparably small computational costs compared to the dynamical models of the atmosphere and the ocean. However, phenomena at the intraseasonal time scale have been difficult to predict with the use of machine learning methods because complex interactions between processes with various spatio-temporal scales that range from the convective to seasonal scales play important role in determining its time evolution. Particularly with regards to the MJO forecasts, machine learning models have been outperformed by dynamical models [25,26].

Here, we employ the reservoir computing method to advance the machine learning prediction of the MJO. The reservoir computing method is a brain-inspired machine-learning technique that constructs a data-driven dynamical model (hereafter reservoir computing models) [27–31]. By training on a time series of macroscopic variables of high-dimensional dynamics, the method can efficiently predict time series and frequency spectra of its chaotic behaviors [32,33]. For example, it is useful for predicting the statistical quantities of a chaotic fluid flow, which cannot be calculated directly from a numerical simulation of the Navier–Stokes equation due to its high computational cost [34]. In this study, we construct a reservoir computing MJO prediction model, trained only by the time series of a macroscopic variable, with a performance competitive with the state-of-the-art physics based dynamical models. Our results demonstrate that the inherent predictability of some MJO cases is longer than have been expected from studies by dynamical models [17].

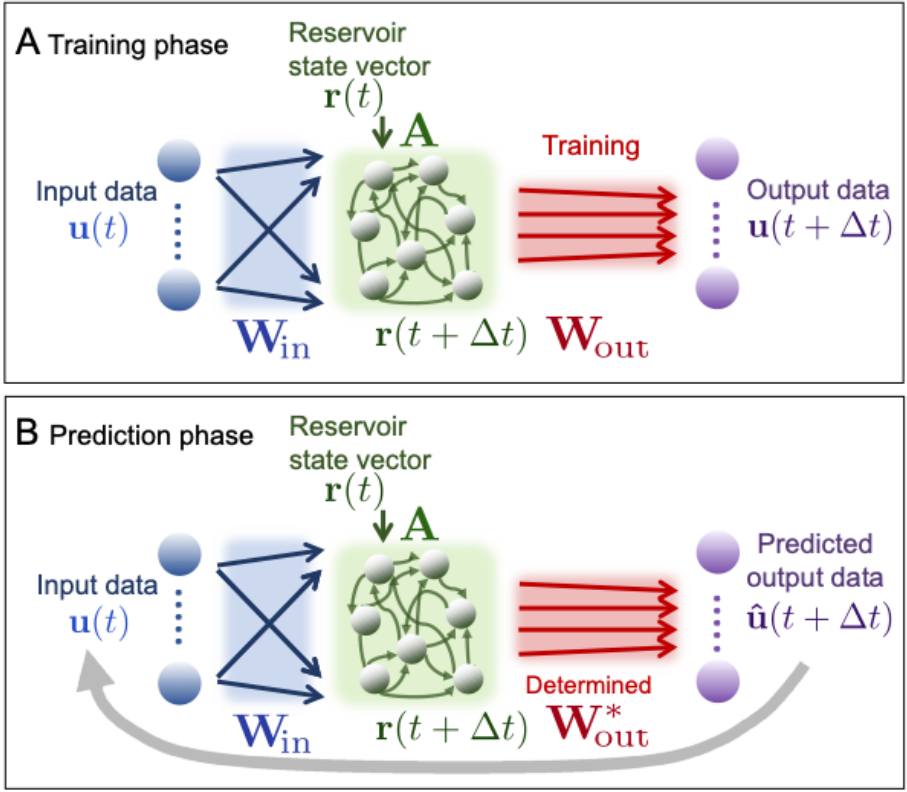

**Fig. 1**. **Schematic picture of reservoir computing.** (**A**) In the training phase, the input data $\boldsymbol{u}(t)$ for time $t$ is fed to the reservoir state vector $\boldsymbol{r}(t)$ through input weight matrix $\boldsymbol{W}_{\text{in}}$ and the output weight matrix $\boldsymbol{W}_{\text{out}}$ is determined by reservoir computing. (**B**) In the prediction phase, the time evolution of $\boldsymbol{u}(t)$ for time $t + \Delta t$ is predicted as $\hat{\boldsymbol{u}}(t + \Delta t)$ from the $\boldsymbol{W}^{*}{}_{\text{out}}$ determined in the training phase.

A reservoir is a recurrent neural network whose internal parameters are adjusted to fit the data in the training process [27,28]. It is trained by feeding an input time series and fitting a linear function of the high dimensional reservoir state vector to the desired output time series (Fig. 1). The construction of a reservoir computing model simply assumes recurrent and deterministic property of the input time series and does not involve any physical knowledge of the input data. The reservoir computing model of this study is described by:

$$\begin{cases} \boldsymbol{r}(t + \Delta t) = (1-\alpha)\,\boldsymbol{r}(t) + \alpha \tanh\big(\boldsymbol{A}\,\boldsymbol{r}(t) + \boldsymbol{W}_{\text{in}}\,\boldsymbol{u}(t)\big) \\ \qquad\quad \hat{\boldsymbol{u}}(t + \Delta t) = \boldsymbol{W}^{*}_{\text{out}}\,\boldsymbol{r}(t + \Delta t) \end{cases} \qquad (1)$$

where $\boldsymbol{u}(t) \in \mathbb{R}^{M}$ is both the input variable vector, $\boldsymbol{r}(t) \in \mathbb{R}^{N}$ ($N \gg M$) is the reservoir state vector, $\boldsymbol{A} \in \mathbb{R}^{N \times N}$, $\boldsymbol{W}_{\text{in}} \in \mathbb{R}^{N \times M}$, and $\boldsymbol{W}^{*}_{\text{out}} \in \mathbb{R}^{M \times N}$ are reservoir, input, and output weight matrices, respectively, $\alpha$ ($0 < \alpha \leq 1$) is the coefficient that adjusts the nonlinearity of the dynamics of $\boldsymbol{r}$, and $\Delta t$ is the time step. We define $\tanh(\mathbf{q}) = (\tanh(\mathrm{q}_1), \tanh(\mathrm{q}_2), \dots \tanh(\mathrm{q}_N))^{T}$, for a vector $\mathbf{q} = (\mathrm{q}_1, \mathrm{q}_2, \dots \mathrm{q}_N)^{T}$, where $T$ represents the transpose of a vector. $\boldsymbol{W}^{*}_{\text{out}}$ is determined to satisfy $\boldsymbol{u}(t) \approx \boldsymbol{W}_{\text{out}}\,\boldsymbol{r}(t)$ using the training data $\boldsymbol{u}(t)$, where $\boldsymbol{W}_{\text{out}}$ is the output weight matrix in the training phase. Further details on the construction of the reservoir computing model are provided in the supplementary materials. In the prediction phase, the predicted variable $\hat{\boldsymbol{u}}(t + \Delta t)$ is obtained from $\boldsymbol{u}(t)$ and $\boldsymbol{r}(t)$, using eqn. (1) with fixed $\boldsymbol{A}$, $\boldsymbol{W}_{\text{in}}$, and $\boldsymbol{W}^{*}_{\text{out}}$. A successful training will give $\hat{\boldsymbol{u}}(t + \Delta t)$ that approximates the desired unmeasured quantity $\boldsymbol{u}(t + \Delta t)$.

The objective of our reservoir computing model is to predict the sequence of the Realtime Multi-variate MJO (RMM) index [6], which is widely accepted as the standard proxy for diagnosing the state of an MJO[1]. It captures the signals of the MJO as an envelope of convective activities coupled to planetary-scale circulation from the leading pair of principal components (RMM1, RMM2) of the equatorial outgoing longwave radiation and zonal winds at 850 hPa and 200 hPa. The RMM calculated from data without smoothing in time [6] has been applied to machine learning prediction of the MJO [25,26]; however, their machine learning predictions were susceptible to degradation ascribed to noises in unsmoothed data from atmospheric variabilities outside of the MJO timescale. Moreover, signals at time scales longer than the MJO needs to be removed from the training data for the machine learning to identify recurring patterns. Thus, to refine the RMM time series to train our reservoir computing model for MJO prediction, signals outside of the MJO frequency range were removed from the raw data by an application of a filter that approximately retains signals only between 20 days and 120 days frequency range [35].

The Lanczos filter [36], which is conventionally used to filter MJO signals, cannot be employed as the filter to generate the training data for the machine learning. This is because the Lanczos filter, whose weights are symmetric in time, requires data from both the past and the future to calculate a filtered value at a certain point in time (Fig. 2 A). To resolve this problem, we design a novel filter, applicable for real-time use, that does not require data from the future. The filter $\Psi_{r^+,r^-,c}$ is defined as:

$$\Psi_{r^+,r^-,c}(t) = F_{r^+,r^-}(t) \frac{\sin(\frac{t}{c} - \pi)}{(\frac{t}{c} - \pi)},$$

where

$$F_{r^+,r^-}(t) = \begin{cases} \dfrac{\sin\left(\frac{t}{r^+}\right)}{t} - \dfrac{\sin\left(\frac{t}{r^-}\right)}{t} & (t \leq 0) \\ 0 & (t > 0) \end{cases},$$

and $c$ is a parameter that adjusts the center of the weights. We set the parameters as $(r^+, r^-, c) = (\frac{20}{\pi}, \frac{120}{\pi}, 14)$ to remove the signals at frequencies lower than 120 days and higher than 20 days. The shape of the filter function in real-space and in Fourier space is compared against that of the Lanczos filter in Fig. 2 A, B. In contrast to the Lanczos filter, the weights of the new filter vanish at $t = 0$ and require only the data from the past. The asymmetric weights of the new filter make it suitable for its application to real-time use such as filtering the input variable data for machine learning predictions. Due to the asymmetry, the center of the weight of the new filter shifts backward only by approximately eight days. Hereafter, this filter will be referred to as the real-time band-pass filter (RB filter). The RMM time series is calculated from data filtered by the RB filter in this study (see methods for details).

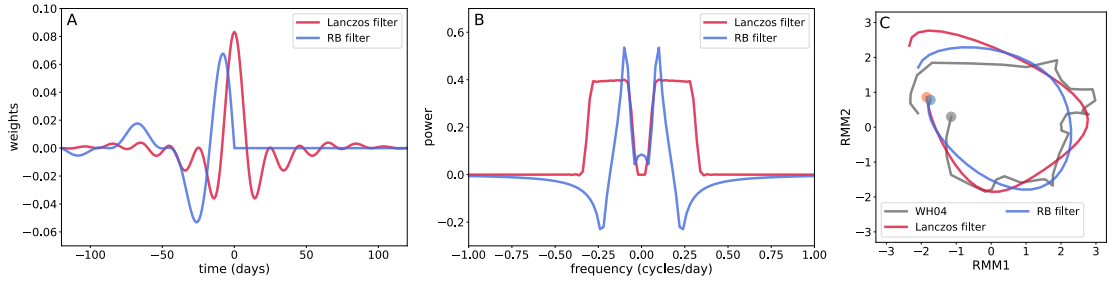

**Fig. 2**. **Comparison of Real-time Band-pass filter and Lanczos filter.** The shape of the Lanczos (red) and RB (blue) filters are shown in (**A**) real space and in (**B**) Fourier space. (**C**) Sample trajectories, from 1st December 2018 (indicated by circles) to 9th January 2019, for the original Wheeler and Hendon 2004 RMM index (WH04, grey), and RMM index filtered by Lanczos (red) and RB (blue) filters and RMM2 replaced by 12-day time-delay coordinate of RMM1. The axis for both RMM2 and 12-day time-delay coordinate of RMM1 is labeled as RMM2 for brevity.

Furthermore, the MJO prediction is refined by employing the RMM phase space spanned by RMM1 and its delay coordinate to diagnose the state of the MJO. That is, we replace RMM2 with the delay coordinate of RMM1 and eliminate the model prediction of RMM2. This enhances the recurrency of the input data and contributes to the robustness of the trained model. The modification is founded on the expectation that RMM2 can be reconstructed from the delay coordinate of RMM1, since RMM1 and RMM2 are orthogonal by definition and the trajectories of the projections of MJO events on the RMM phase space are near circular. The delay time of the delay coordinate is chosen at 12 days, when the auto-correlation of RMM1 crosses zero for the first time. The correlation coefficient of RMM2 and 12-day delay coordinate variable of RMM1 is 0.75. The trajectories of the RB-filtered and Lanczos filtered RMM sequences with RMM2 replaced by the 12-day delay coordinate of RMM1 is compared with the original Wheeler and Hendon RMM (WH04) [6] in Fig. 2C. We confirm that the RB filter removes signals from slow variabilities and noises as effectively as the Lanczos filter and that the trajectory of the RMM sequence on the phase space with RMM2 replaced by the 12-day delay coordinate of RMM1 is similar to that of the WH04 RMM on phase space spanned by RMM1 and RMM2. Thus, we focus on the RMM phase space spanned by RMM1 and its 12-day delay coordinate, which we will refer to as the machine learning RMM (ML-RMM) phase space. The relevance of ML-RMM phase space to the conventional one spanned by RMM1 and RMM2 is further discussed in the supplementary materials (Fig. S1). We will denote RMM1 and its time-delay coordinate at time $t$ as $a(t) \coloneqq \mathrm{RMM1}(t)$ and $b(t) \colon= a(t - 12)$.

It is known that a chaotic attractor can be reconstructed by some observable variables and its delay coordinates [37,38]. For the construction of a reservoir computing model, it is efficient to take the delay coordinate variable with an appropriate delay time as the input when the number of observable variable is smaller than the effective dimension of the attractor [33]. A suitable delay time and dimension of the delay coordinate of RMM1 is inferred by computing its auto-correlation function. Thus, an $M$-dimensional delay coordinate vector of RMM1 is introduced as the input and output variable vector $\boldsymbol{u}$ in Eq. (1):

$$\boldsymbol{u}(t) = \big(\mathrm{RMM1}(t), \mathrm{RMM1}(t - 1\Delta\,\tau), \dots, \mathrm{RMM1}(t - (M-1)\Delta\,\tau)\big),$$

where $\Delta\tau$ is the delay time, and $(\Delta\,\tau, M) = (6, 7)$. The pair of parameters are chosen so that the behavior of $\boldsymbol{u}(t)$ is deterministic and has recurrency, which are essential properties for successful modelling. The reservoir model (Eq. (1)) of the RMM1 time sequence is constructed by determining $\boldsymbol{W}^{*}{}_{\mathrm{out}}$. The time series of the RMM1 data from 30th December 1986 to 29th December 2011 was used as the training data. The optimal reservoir computing model was selected from evaluation of test cases of RMM1 forecasts initialized from every day between 8th April 2014 and 6th July 2014. The selected model is used throughout this study for all predictions.

The predicted variables at time $t$ initialized from time $p$ are denoted as $\hat{a}(t, p)$ and $\hat{b}(t, p)$. We note that $\hat{b}(t, p)$ is predicted by the reservoir computing model simultaneously with $\hat{a}(t, p)$. The relationship $\hat{b}(t, p) = \hat{a}(t - 12, p)$ would hold only in an ideal case in which the model learns the delay coordinate of the RMM1 perfectly. The reference time series in this case are $a(t)$ and $b(t)$. We compare the time series of predicted variables against the reference time series using the bivariate correlation coefficient (COR) [16,39], defined by the equation:

$$\text{COR}(q) := \frac{\sum_{p=1}^{N} \left( a(p+q)\hat{a}(p+q,p) + b(p+q)\hat{b}(p+q,p) \right)}{\sqrt{\sum_{p=1}^{N}((a(p+q))^2 + \left(b(p+q)\right)^2)} + \sqrt{\sum_{p=1}^{N}((\hat{a}(p+q,p))^2 + (\hat{b}(p+q,p))^2)}} ,$$

where $q$ is the forecast lead time. The COR corresponds to a covariance between the actual vector $(a(t), b(t))$ and the predicted vector $(\hat{a}(t,p), \hat{b}(t,p))$, and is conventionally used to evaluate the MJO prediction skills of dynamical and statistical models [40,41]. Here, the $N = 2010$ is the number of predictions initialized for all days between 28th July 2014 and 28th January 2020. In Fig. 3, we show the time series of $\text{COR}(q)$ for all predictions and three cases, the details of which will be described next. The $\text{COR}(q)$ stays above 0.5 for 28 days for all predictions. The threshold value 0.5 is customarily adopted for MJO prediction skill score [16]. This signifies that the expectancy of the skill score of the model is at three weeks for all days, including periods devoid of MJO activity. The forecast skill was evaluated as three weeks in consideration of the approximate 8-day shift by the RB filter as discussed above.

It is customary to evaluate the skill of MJO predictions from the forecasts of periods when MJO events are identified [14]. We reevaluate the forecast skill of the reservoir model following the custom. Here, the MJO events were identified as continuous sequences from phase 2 to phase 7 on the RMM phase space spanned by RMM1 and its delay coordinate of 12 days [8,35] (See methods for details). For the predictions initialized on three, five, and seven days before the onsets of MJO events, the COR remains above 0.5 for 38 days for all three cases. Considering the 8-day shift by the RB filter, this signifies that the constructed model has the potential to skillfully forecast the time evolutions of the MJO events for 30 days. Counterintuitively, the COR decays below 0.5 faster for the forecasts initialized three days before the MJO onsets than those for five and seven days before the MJO onsets. We note however, that this is consistent with the fact that the predictions reach the terminations of MJO events faster for predictions that are initiated closer to the onsets.

The performance of the MJO prediction on individual cases are examined to illuminate the similarity between the predicted and the actual trajectories of the RMM. Figure 4 compares the actual and predicted trajectories on the ML-RMM phase, prediction errors, and the phase difference for the 10th (A, B, C), 26th (D, E, F), and 50th (G, H, I) best performing cases in terms of mean error over the first 60 days of the prediction. The three samples are chosen so that there are no overlaps in the forecast lead times. The errors are measured by the distance between the actual $(a(t), b(t))$ and the predicted $(\hat{a}(t, p), \hat{b}(t, p))$ vectors. The phase difference is evaluated from the cosine of the angle between $(a(t), b(t))$ and $(\hat{a}(t, p), \hat{b}(t, p))$ $(\cos(\theta_p(t)))$. In all three cases, the error remains below 1.4, the threshold of the root mean square errors of the predicted RMM adopted to evaluate the skills of climate simulations [42], well beyond two months ($>$ 75 days). The prediction also stays in phase $(\cos(\theta_p(t)) > 0.7)$ for nearly two months (58, 83, and 76 days for the 10th, 26th, and 50th best case, respectively). We note that the rapid increases in phase differences occur when the amplitude of the RMM1 decreases. This is reasonable considering that the RMM phases become physically meaningless with diminishment of its amplitude. These results indicate that our reservoir computing model can predict the state of some MJO events well beyond the estimated inherent predictability limit of 7 weeks from dynamical model studies [17]. This inference is supported by cases of RMM1 predictions that skillfully forecast RMM1 phases for longer than 120 days (see Fig. S2).

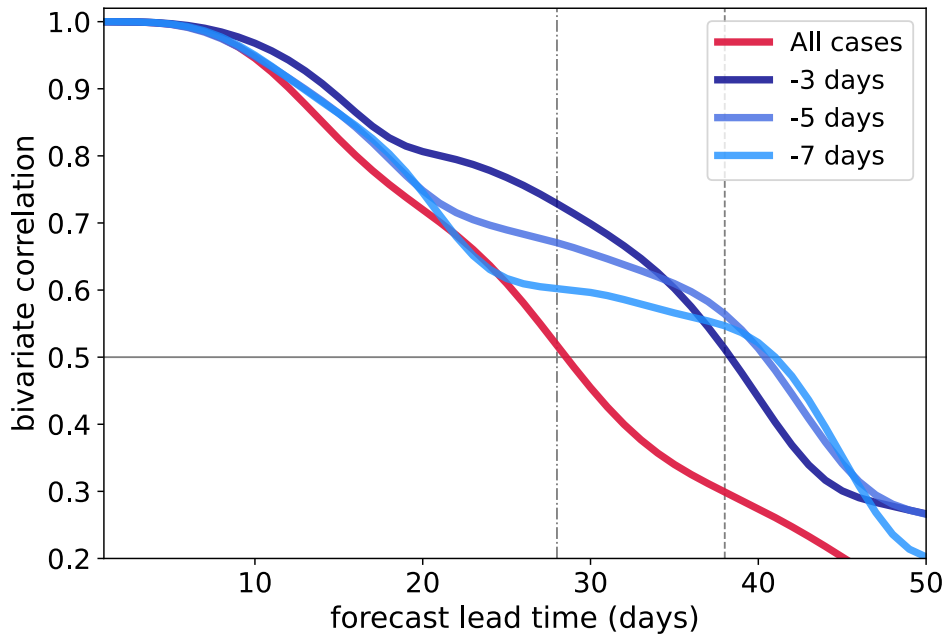

**Fig. 3**. **Bivariate correlation coefficient.** The mean bivariate correlation coefficient (COR) as a function of forecast lead time (days) for all 2010 predictions (red), and for predictions initialized on 3 (navy), 5 (blue), and 7 (light blue) days before onsets of MJO events. The dash-dot and the dotted lines indicate 28th and the 38th day in the forecast lead time, respectively.

We constructed a computationally inexpensive machine learning model, using the reservoir computing technique, that is capable of month-long forecasts of the state of the MJO. This prediction skill is superior to that of preceding machine learning methods and is matched only by physics-based dynamical models that inevitably demand the state-of-the-art supercomputers [13,14,40,43]. It is remarkable that our model was trained only by the macroscopic time series of the RMM1. This signifies that intricate information of the atmospheric and oceanic states that influences the MJO [44,45] were implicitly incorporated into the reservoir state variables of the neural network. The extended prediction skill of our reservoir model is attributed to the refinement of the training data. The signals from slow variability and high frequency noise were filtered out from the input data with the RB filter to restrict the degrees of freedom of the training. This was essential because it was necessary for the model to efficiently learn from merely 26 years of RMM1 data with a limited number ($<100$) of MJO events. To further enhance the efficacy of the reservoir computing, we introduced the delay coordinate variable of RMM1 to employ suitably correlated variables as our training data [33]. It is of interest how the extension of the training data with accumulation of observational data in the future will enhance the performance of the reservoir model.

The best performing forecasts by our reservoir model predicted the RMM time series for more than two months. These results indicate that some MJO events are inherently predictable beyond the potential predictability estimates made from dynamical model studies at seven weeks [17]. This implies a possibility for significant improvements in dynamical models to extend their lead time in MJO prediction, which is crucial for reliable global weather forecasts. However, observations suggest that global warming alters the characteristics of the MJO [8], meaning that the applicability of machine learning models trained on historical data for MJO predictions could be undermined by climate change in the future. Furthermore, the reservoir model of this study can only forecast the RMM sequence and cannot directly assess the impact of the MJO on the midlatitude weather. Thus, dynamical models are expected to continue to be an imperative tool for predicting and understanding the behaviors of our atmosphere and it is important to make the efforts to exploit machine learning weather predictions to advance the dynamical models.

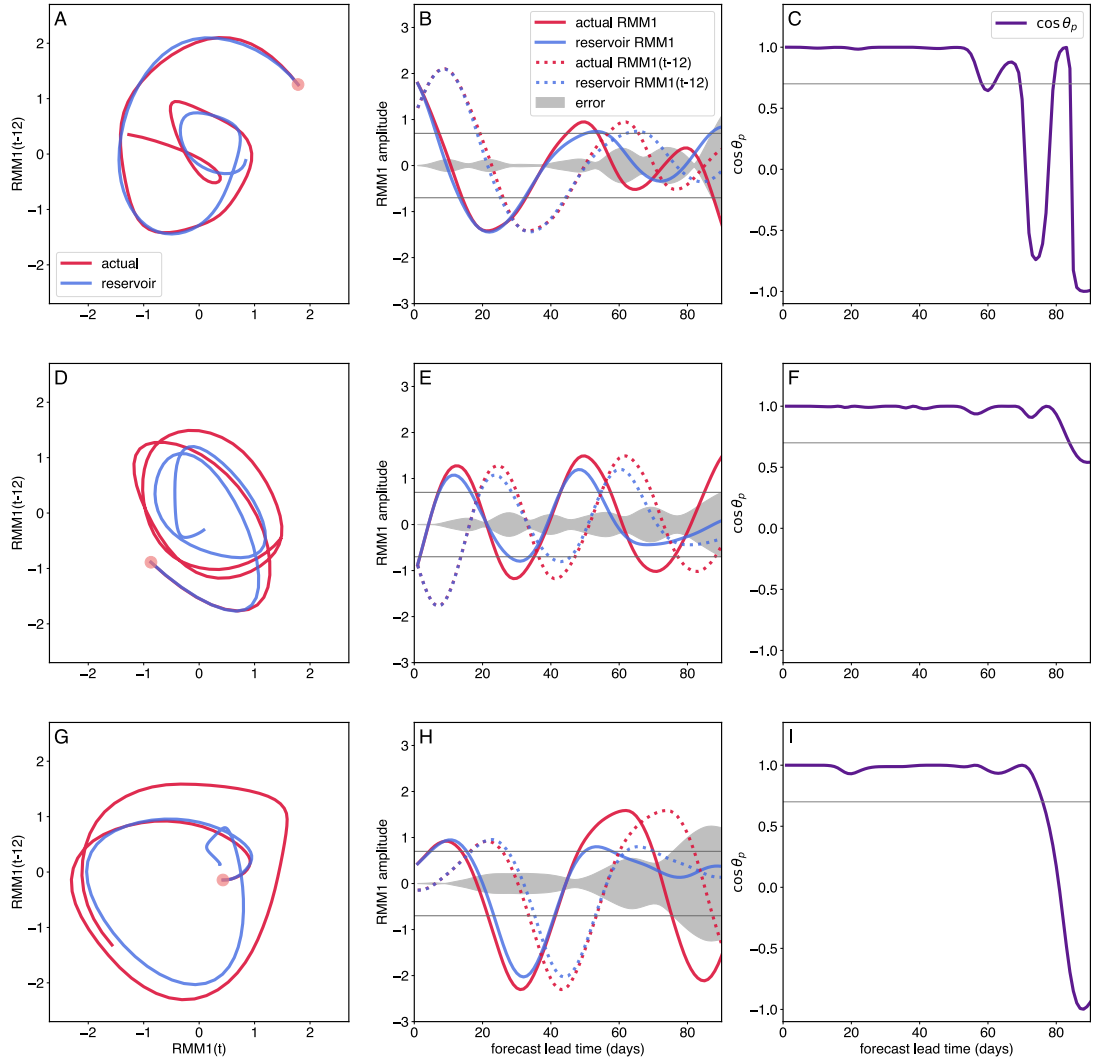

**Fig. 4. Samples of best performing cases of RMM1 predictions and their errors.** The (**A**, **B**, **C**) 10th, (**D**, **E**, **F**) 26th, and (**G**, **H**, **I**) 50th best performing cases of RMM1 predictions initialized from 19th June 2019, 14th April 2018, and 9th December 2015 (indicated by the red dots), respectively. (**A**, **D**, **G**) The trajectories of the actual (red) and the predicted (blue) RMM1 ($a(t)$ and $\hat{a}(t,p)$) and its time-delay coordinate ($b(t,p)$ and $\hat{b}(t,p)$) are shown on the RMM phase space and (**B**, **E**, **H**) as a function of the forecast lead time with the errors shown as the width of the gray shade. (**C**, **F**, **I**) The time series of prediction errors measured by the cosines of the angles between $(a(t), b(t))$ and $(\hat{a}(t,p), \hat{b}(t,p))$ ($\cos(\theta_p(t))$ ). The gray lines at ±0.7 in panels **B**, **E**, **H** indicate the threshold for the error = 1.4 and $\cos(\theta_p(t)) = 0.7$ in panels **C**, **F**, **I**.

## Acknowledgments

**Funding:**

Japan Society for the Promotion of Science KAKENHI Grants, 21K13991 and 20H05730 (TS)

Japan Society for the Promotion of Science KAKENHI Grants, 22K17965 (KN)

Japan Science and Technology Agency PRESTO, 22724051(KN)

Japan Society for the Promotion of Science KAKENHI Grants, 20H01819 (TY)

Japan Society for the Promotion of Science KAKENHI Grants, 20H05728 and 22H01297 (DT)

MEXT Program for Promoting Researches on the Supercomputer Fugaku, hp210166 and hp220167 (DT)

Japan Society for the Promotion of Science KAKENHI Grants, 20J11246 (TJ)

Japan Society for the Promotion of Science KAKENHI Grants, 19KK0067 and 21K18584 (YS)

"Joint Usage/Research Center for Interdisciplinary Large-scale Information Infrastructures" in Japan, jh210027and jh220007, (YS)

HPCI System Research Project, hp210072 (YS)

Japan Society for the Promotion of Science KAKENHI Grants, 16H04048 and 20H05729 (HM)

**Author contributions:**

Conceptualization: KN, TS, DT, TY, HM, YS

Methodology: KN, TY, TS, HM, YS

Investigation: KN, TS, DT, HM, YS

Visualization: TS, KN

Funding acquisition:  TS, HM, KN, DT, TJ, TY, YS

Supervision: HM, YS

Writing – original draft: TS, KN, HM, YS

Writing – review and editing: TS, HM, KN, DT, TJ, TY, YS

**Competing interests:** Authors declare that they have no competing interests.

**Data availability**

NOAA-OLR data are available at https://www.psl.noaa.gov/data/gridded/data.olrcdr.interp.html .

NCEP-DOE reanalysis data for zonal wind data are available at

https://psl.noaa.gov/data/gridded/data.ncep.reanalysis2.html .

The original Wheeler and Hendon 2004 RMM time series are available at

http://www.bom.gov.au/climate/mjo/ .

**Code availability**

All source codes of our reservoir model, filter-function of the RB filter, input and output data of the reservoir computing, and the list of MJO events will be provided via zenodo before publication of this work.

<div align="center">

**References**

</div>

**Methods**

**Reservoir computing technique**

The method of determining the output weight matrix $\boldsymbol{W}^*{}_{\text{out}}$ of our reservoir computing machine learning model is described. The time development of the reservoir state vector $\boldsymbol{r}(l\,\Delta\,t)$ is determined by:

$$\boldsymbol{r}(t + \Delta\,t) = (1 - \alpha)\,\boldsymbol{r}(t) + \alpha \tanh\big(\boldsymbol{A}\,\boldsymbol{r}\,(t) + \boldsymbol{W}_{\text{in}}\,\boldsymbol{u}(t)\big)\,, \qquad (1 - \text{M})$$

where $\{\boldsymbol{u}(l\,\Delta\,t)\}\,(-L_0 \leq l \leq L)$ is the training time series data, $L_0$ is the transient time, and $L$ is the time length to determine $\boldsymbol{W}^*{}_{\text{out}}$. For given random matrices $\boldsymbol{A}$ and $\boldsymbol{W}_{\text{in}}$, we determine $\boldsymbol{W}_{\text{out}}$ so that the following quadratic form takes the minimum:

$$\sum_{l=0}^{L} \left\|\boldsymbol{W}_{\text{out}}\,\boldsymbol{r}\,(l\Delta t) - \boldsymbol{u}\big((l+1)\Delta t\big)\right\|^2 + \beta[Tr\,(\boldsymbol{W}_{\text{out}}\,\boldsymbol{W}_{\text{out}}^T\,)], \qquad (2 - \text{M})$$

where $\|\boldsymbol{q}\|^2 = \boldsymbol{q}^T\boldsymbol{q}$ for a vector $\boldsymbol{q}$. The minimizer is

$$\boldsymbol{W}^*{}_{\text{out}} = \delta\boldsymbol{U}\delta\boldsymbol{R}^T(\delta\boldsymbol{R}\delta\boldsymbol{R}^T + \beta\,\boldsymbol{I})^{-1} \qquad (3 - \text{M})$$

where $\boldsymbol{I}$ is the $N \times N$ identity matrix, $\delta\boldsymbol{R}$ and $\delta\boldsymbol{U}$ are the matrices whose $l$-th column is $\boldsymbol{r}\,(l\Delta t)$ and $\boldsymbol{u}\,(l\Delta t)$, respectively. (see Lukosevivcius and Jaeger (2009)[46] P.140 and Tikhonov and Arsenin (1977)[47] Chapter 1 for details).

Note that $\boldsymbol{A}$ is chosen to have a maximum eigenvalue $\rho\,(|\rho| < 1)$ in order for eqn. (2-M) to satisfy the so called echo state property. It is known that addition of noises to the training time series data is potentially useful in the construction of a data-driven model [22]. Further details on the reservoir computing can be found in preceding literatures [32–34,48].

The set of parameter values used to construct the reservoir computing model is shown in Table 1. We determine $\boldsymbol{W}_{\text{out}}$ using the training time series data $\boldsymbol{u}$, which in this case is the RMM1 data from 30th December 1986 ($t = 0$) to 29th December 2011 ($t = 9131$). The optimal reservoir computing model was selected based on predictions of the RMM1 initialized every day between 8$^{th}$ April 2014 and 16$^{th}$ July 2014 by using $\boldsymbol{W}_{out}$ for a given $\boldsymbol{A}$ and $\boldsymbol{W}_{in}$. We selected a model with the smallest prediction error $\max_{t\in[1,5]}|u_1(t)-\hat{u}_1(t)|$ and $\max_{t\in[1,90]}|u_1(t)-\hat{u}_1(t)|$, where $u_1(t)$ is the first component of $\boldsymbol{u}$ and $\hat{u}_1(t)$ is the predicted variables of $u_1(t)$.

| | parameter | value |
|---|---|---|
| $M$ | Dimension of input and output variables | 7 |
| $N$ | Dimension of reservoir state vector | 1000 |
| $\Delta t$ | Time step of the model | 1 (day) |
| $\rho$ | Maximal eigenvalue of $\boldsymbol{A}$ | 0.8 |
| $\alpha$ | Nonlinearity degree in the model | 0.7 |
| $\beta$ | Regularization parameter | 0.01 |
| $\Delta\tau$ | Delay-time for input and output variables | 6 (day) |

**Table 1. The list of parameters and their values for the selected reservoir computing model**

**MJO detection method**

The RMM is calculated from the combined empirical orthogonal functions of the outgoing longwave radiation data from National Oceanic and Atmospheric Administration [49], and zonal wind data from National Centers for Environmental Prediction-Department of Energy reanalysis [50]. With the exception of replacing RMM2 with the 12-day time delay coordinate of RMM1, the orientation of RMM1 and definitions of the RMM phases follow the convention introduced by Wheeler and Hendon [6]. The MJO events were identified from time sequences that were projected on to the RMM index from phase 2 to phase 7, while satisfying the following four conditions employed by Suematsu and Miura (2018) [35]: (1) Phases do not skip forward nor recede backward by more than one phase. (2) The average amplitude is greater than the critical value of 0.8. (3) Period of consecutive days with amplitude below 0.8 is less than 15. (4) Transition from phase 2 to phase 7 takes 20 to 90 days.

**Supplementary materials**

**Validity of the Machine Learning-RMM Phase Space**

The relevance of employing the RMM phase space spanned by RMM1 and its delay coordinate, the machine learning RMM (ML-RMM) phase space, to describe the MJO instead of that spanned by RMM1 and RMM2 is discussed. Conventionally, MJO events are identified using RMM phase space spanned by the first two orthogonal functions, RMM1 and RMM2, of 20 - 120 day Lanczos bandpass filtered [36] outgoing longwave radiation and zonal winds at 850hPa and 200 hPa. Figure S1 compares the composites of 1979 – 2020 December to February outgoing longwave radiation for each of the RMM phases spanned by the conventional RMM1 and RMM2 with ML-RMM phase space.

The composites indicate that the definition of the ML-RMM phases (Fig. S1 B) can capture the characteristic of the MJO convection to shift eastward from the Indian Ocean to the Western Pacific as the conventional method (Fig. S1 A). We note however, that compared to the conventional method, the convective signals over the Indian Ocean in the ML-RMM phase 2 is weaker. This may be a caveat to our method that arises from replacing the RMM2 with the delay coordinate of RMM1, since the structure of the eigenvector of RMM2 reflects the state of the atmosphere with deep convection over the Indian Ocean (see Fig. 1 in 30). Despite the abovementioned concern, the method employed in this study is capable of adequately tracking MJO events on the RMM phase space (Fig. 2C).

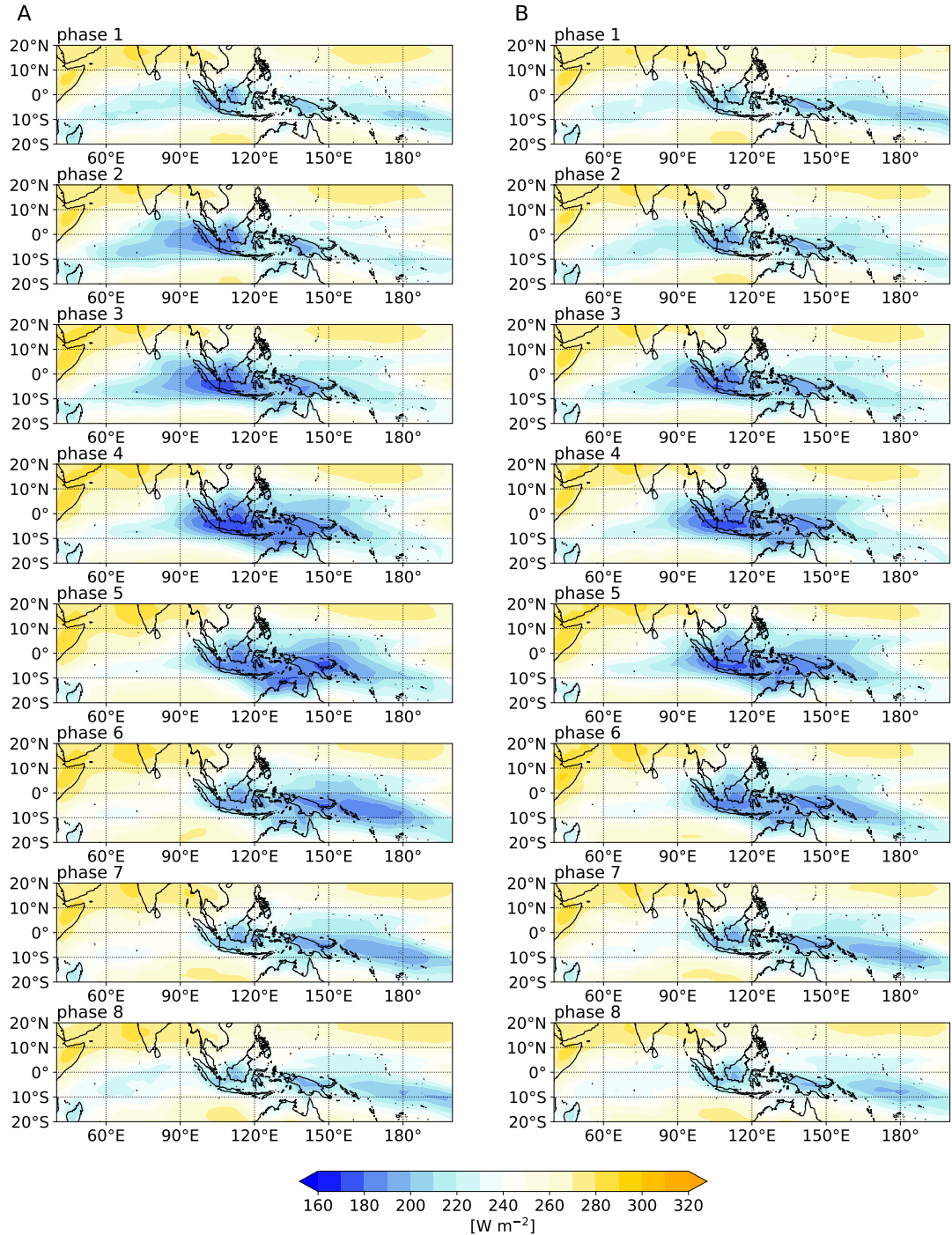

**Fig. S1.** Composites of 1979 – 2020 December to February outgoing longwave radiation for each of the RMM phases on the (A) conventional RMM1 and RMM2 phase space calculated from 20-120 days Lanczos bandpass filtered data and (B) on the ML-RMM calculated from the 20-120 days RB filtered data.

**Examples of best performing cases in terms of phase prediction**

The best performing prediction cases in terms of ML-RMM phase predictions are examined. Figure S2 shows the best three cases evaluated by the first day the cosine of the phase difference between the actual $(a(t), b(t))$ and the predicted $(\hat{a}(t, p), \hat{b}(t, p))$ vector, $\cos(\theta_p(t))$, becomes less than 0.7. The first (Fig. S2. **A**, **D**), second (Fig. S2. **B**, **E**) and third (Fig. S2.**C**, **F**) best performing cases are the predictions of ML-RMM initiated on 10th October 2017, 26th March 2019, and 18th April 2018, respectively. In all three cases, the predictions stay in phase $(\cos(\theta_p(t)) > 0.7)$ for longer than 120 days. However, there is a tendency for the amplitudes to be underestimated in these cases, which leads to growth in error as measured by the distance between $(a(t), b(t))$ and $(\hat{a}(t, p), \hat{b}(t, p))$ from early stages of the predictions. Thus, while the long predictability of the RMM phases over 120 days suggest the possibility of predicting the MJO over a season (three months), overcoming the difficulty of accurately predicting the RMM phase and amplitude simultaneously remains a challenge.

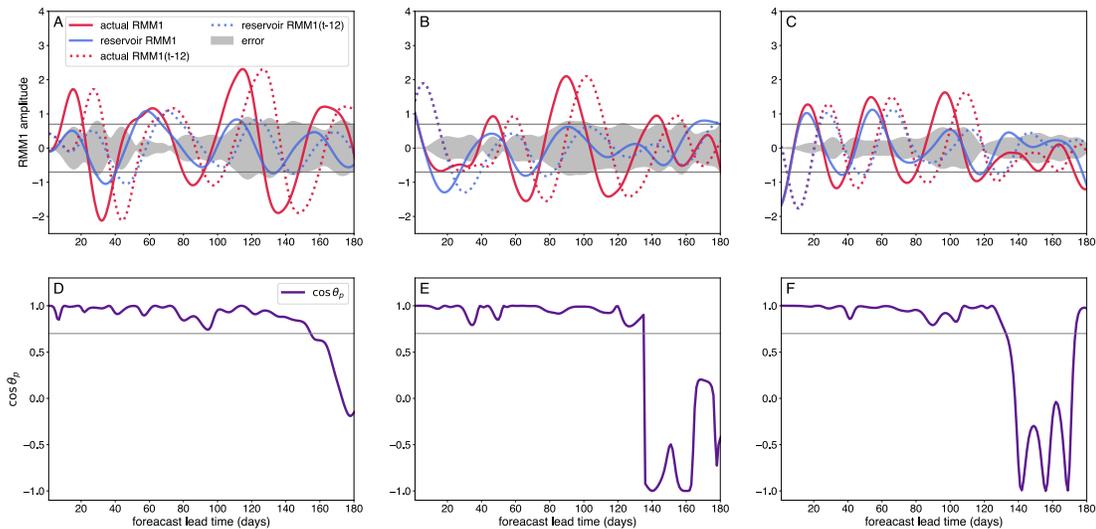

**Fig. S2. The three best performing prediction of ML-RMM time series in terms of phase prediction.** (**A**, **B**, **C**) Predictions of ML-RMM initialized from (**A**, **D**) 2nd October 2017, (**B**, **E**) 18th March 2019, and (**C**, **F**) 10th April 2018, which are the three best prediction cases measured by the cosine of the phase difference between the actual $(a(t), b(t))$ and the predicted vectors $(\cos(\theta_p(t)))$.

(**D**, **E**, **F**) show the time evolution of the $\cos(\theta_p(t))$. The width of the grey shades in A, B, C indicates the error measured by the distance between $(a(t), b(t))$ and $(\hat{a}(t, p), \hat{b}(t, p))$.